\documentclass[12pt,a4paper]{article}
\usepackage{feynmf}  
\renewcommand{\thefootnote}{\fnsymbol{footnote}}
\newcounter{line}

\def\ibid#1#2#3{{\it ibid.}{\bf #1} (#2) #3}
\def\ie{\hbox{\it i.e.}{}}

\def\etc{\hbox{\it etc}{}}
\def\nn{\hspace{2mm}}
\def\sss{\scriptscriptstyle}
\newcommand{\MeV}{\mbox{\rm MeV}}
\newcommand{\GeV}{\mbox{\rm GeV}}
\newcommand{\eV}{\mbox{\rm eV}}
\def\sleq{\raisebox{-.6ex}{${\textstyle\stackrel{<}{\sim}}$}}
\def\sgeq{\raisebox{-.6ex}{${\textstyle\stackrel{>}{\sim}}$}}
\def\Tr{{\rm Tr}{}}

\def\Bar#1{\overline{#1}}

\def\ket#1{\left| #1\right\rangle}

\def\sVEV#1{\left\langle #1\right\rangle}

\def\abs#1{\left| #1\right|} 
\def\cL{{\cal L}}
\def\AGUT{{}\;\;\raisebox{.9ex}{$\times$}\raisebox{-.5ex}%
{$\!\!\!\!\!\!\!\!\sss i=1,2,3$} \,(SMG_i \times U(1)_{\sss B-L,i})}%
\begin{document}
\begin{titlepage}
%\hfill
%\title{ 
\begin{flushleft}
\vspace{-1.5cm}
\vbox{
    \halign{#\hfil        \cr
           DESY 02-006    \cr
           GUTPA/01/12/01 \cr
           NBI-HE-01-12    \cr
           hep-ph/0201152 \cr
           January 2002 \cr
           } % end of \halign
      }  % end of 
\end{flushleft}
\vspace*{-1cm}
%\vspace*{-2mm}
\begin{center}
{\Large {\bf {}Family replicated gauge groups and 
large mixing angle solar neutrino solution}\\}

\vspace*{7mm}
{\ C. D. Froggatt}$^{\it a,}$\footnote[1]{E-mail: c.froggatt@physics.gla.ac.uk},
{\ H. B. Nielsen}$^{\it b,c,}$\footnote[2]{E-mail: hbech@mail.desy.de; hbech@nbi.dk}
and {\ Y. Takanishi}$^{\it b,c,}$\footnote[3]{E-mail: yasutaka@mail.desy.de; yasutaka@nbi.dk}
            
\vspace*{.2cm}
{\it $^a$ Department of Physics and Astronomy,\\
Glasgow University,\\
Glasgow G12 8QQ, Scotland}\\
\vskip .3cm
{\it $^b$ Deutsches Elektronen-Synchrotron DESY, \\
Notkestra{\ss}e 85,\\
D-22603 Hamburg, Germany}\\
\vskip .3cm
{\it $^c$ The Niels Bohr Institute,\\
Blegdamsvej 17, DK-2100 Copenhagen {\O}, Denmark}\\
\vspace*{.3cm}
%\maketitle
\end{center}
\begin{abstract}
%%%%%%%%%%%%%%%%%%%%%%%%%%%%%%%%%%%%%%%%%%%%%%%%%%%%%%%
We present a modification of our previous family replicated 
gauge group model, which now generates the Large Mixing Angle MSW 
solution rather than the experimentally disfavoured Small Mixing 
Angle MSW solution to the solar neutrino oscillation problem. 
The model is based on each family of quarks and leptons 
having its own set of gauge fields, each containing a replica of 
the Standard Model gauge fields plus a $(B-L)$-coupled gauge 
field. By a careful choice of the Higgs field gauge quantum numbers, 
we avoid our previous prediction that the solar neutrino mixing angle 
is equal order of magnitudewise to the Cabibbo angle, replacing it 
and the well-known Fritzsch relation with the relation 
$\theta_{c}\sim (\theta_{\odot})^{-1/3}\,(m_d/m_s)^{2/3}$. 
At the same time we retain a phenomenologically successful 
structure for the charged quark and lepton mass matrices. 
A fit of all the seventeen quark-lepton mass and mixing 
angle observables, using just six new Higgs field vacuum 
expectation values, agrees with the experimental data 
within the theoretically expected uncertainty of about 
$64\%$, $\ie$ it fits perfectly order of magnitudewise.     

%%%%%%%%%%%%%%%%%%%%%%%%%%%%%%%%%%%%%%%%%%%%%%%%%%%%%%%
\vskip 5.5mm \noindent\
PACS numbers: 12.10.Dm, 12.15.Ff, 14.60.Pq, 14.60.St\\
\vskip -3mm \noindent\
Keywords: Fermion masses, Neutrino oscillations, See-saw mechanism
\end{abstract}

\end{titlepage}
%%%%%%%%%%%%%%%%%%%%%%%%%%%%%%%%%%%%%%%%%%%%%%%%%%%%%%%
\newpage
\renewcommand{\thefootnote}{\arabic{footnote}}
\setcounter{footnote}{0}
\setcounter{page}{2}
%%%%%%%%%%%%%%%%%%%%%%%%%%%%%%%%%%%%%%%%%%%%%%%%%%%%%%%
\section{Introduction}
\indent

The first results on the charge current interactions 
from the Sudbury Neutrino Observatory (SNO) 
collaboration~\cite{SNO} have 
provided an important signal confirming the 
existence of the solar neutrino anomaly
puzzle~\cite{chlorine,sage,gallex,gno,SK8B}: SNO 
detected a flux of non-electron neutrinos,
$\nu_\mu$ and $\nu_\tau$, among solar neutrinos after
travelling from the core of the Sun to the Earth. 
Combination of the SNO results with previous measurements from 
other experiments reveals a confirmation of the standard solar 
model~\cite{Bahcall}, whose predictions of the total 
flux of active $^8$B neutrinos in the Sun agree with the SNO and 
Super-Kamiokande~\cite{SK8B} data.
{}Furthermore, the measurement of the $^8$B and $hep$ 
solar neutrino fluxes shows no significant energy dependence 
of the electron neutrino survival probability in the
Super-Kamiokande and SNO energy ranges. 
These results support the Large Mixing Angle MSW~\cite{MSW} solution 
(LMA-MSW) rather than the Small Mixing Angle MSW solution (SMA-MSW)
to the solar neutrino problem. 

Another important result on the solar neutrino problem, 
reported by the Super-Kamiokande collaboration~\cite{SKDN}, 
is that the day-night asymmetry data disfavour the 
SMA-MSW solution at the $95\%$ C.L.. 
In fact, global analyses~\cite{fogli,cc1,goswami,smirnov} of solar 
neutrino data, including the first SNO results and the day-night 
effect, have confirmed that the LMA-MSW solution gives the best 
fit to the data and that the SMA-MSW solution is very strongly
disfavoured and only accepted at the $3\sigma$ level. The best fit
values of the mass squared difference and mixing angle parameters 
in the two flavour LMA-MSW solution\footnote{The best fit 
parameter values for the LMA solution depend somewhat on the 
analysis method. However they do not change drastically from one 
two flavour analysis to another. We discuss the three flavour analyses 
in section \ref{sec:chooz}.} are 
$\Delta m^2_\odot\approx4.5\times 10^{-5}~\eV^2$ 
and $\tan^2\theta_{\odot}\approx0.35$. 

We have previously attempted to fit all the fermion -- quark 
and lepton -- masses and mixing angles including 
baryogenesis~\cite{NT1,NT2,NT3} in a rather specific model without
supersymmetry or grand unification. The model has the 
maximum number of gauge fields consistent with maintaining 
the irreduciblity of the usual Standard Model fermion 
representations, including three right-handed neutrinos. 
The predictions of this previous model are in order 
of magnitude agreement with all existing experimental data; 
however, only provided we use the SMA-MSW solution. But, for 
the reasons given above, the SMA-MSW solution is now 
disfavoured phenomenologically. So, in this article, we 
present a modified version of the previous model, which 
manages to accommodate the LMA-MSW solution for solar neutrino 
oscillations:  all the fermion mass and mixing angle parameters 
are fitted within a factor of two, using 6 adjustable parameters.

This article is organised as follows: in the next section, we 
define our notation for the charged fermion Yukawa coupling 
matrices, mass matrices and mixing angles. Then, 
in section $3$ we review the family replicated gauge model. In 
section $4$ we discuss the reasons for the modification of our 
model and the introduction of new Higgs fields. The calculation 
is described in section $5$ and the results are presented 
in section $6$. Finally, section $7$ contains 
our conclusion.

\section{Charged fermion masses and their mixing angles}
\indent

In the Standard Model all fermions (apart from the neutrinos)
get a mass via the electroweak spontaneous symmetry breaking -- the 
Higgs mechanism. In extensions of the Standard Model containing 
right-handed neutrinos, the physical light neutrinos get a mass 
via the see-saw mechanism (see discussion of the see-saw mechanism 
in section~\ref{sec:seesaw}). 
The Higgs mechanism generates charged fermion mass terms from their 
Yukawa couplings in the Standard Model Lagrangian: 
\begin{equation}
-{\cal L}_{{\rm charged-fermion-mass}} = 
\overline{Q_{\sss L}}{}Y_{\sss U} \tilde\Phi_{\sss WS}{}U_{\sss R} +
        \overline{Q_{\sss L}}{}Y_D\Phi_{\sss WS}{}D_{\sss R} +
        \overline{L_{\sss L}}{}Y_{\sss E}\Phi_{\sss WS}{}E_{\sss R} + h.c.
\label{L_Higgs}
\end{equation}
Here $\Phi_{WS}$ is the Weinberg-Salam Higgs field,
$Q_L$ denotes the three $SU(2)$ doublets of
left-handed quarks, $U_R$ denotes the three singlets of right-handed 
up-type quarks and $Y_U$ is the three-by-three Yukawa coupling 
matrix for the up-type quarks. Similarly $Y_D$ and $Y_E$ are 
the Yukawa coupling matrices for the down-type quarks and charged 
leptons respectively. 
The $SU(2)$ doublets $\Phi_{WS}$ and $Q_L$ can be represented
as $2$ component column vectors and we then define:
\begin{equation}
\tilde\Phi_{\sss WS} = \left ( \begin{array}{cc} 0 & 1 \\ 1 & 0 
\end{array}\right ) \Phi_{\sss WS}^{\dagger} 
\end{equation}
and
\begin{equation}
\overline{Q_{\sss L}} =
        \overline{ \left ( \begin{array}{c} U_{\sss L} \\
D_{\sss L} \end{array} \right ) }
        = ( \overline{U_{\sss L}} \; \overline{D_{\sss L}} )
\end{equation}
where $\overline{U_{\sss L}}$ are the $CP$ conjugates of the three 
left-handed up-type quarks. After electroweak symmetry breaking 
the Weinberg-Salam Higgs field gets a
vacuum expectation value (VEV) and we obtain the following 
mass terms in the Lagrangian:
\begin{equation}
-{\cal L}_{{\rm charged-fermion-mass}} = \overline{U_{\sss L}} 
\,M_{\sss U}\, 
U_{\sss R} + \overline{D_{\sss L}} \,M_{\sss D}\, D_{\sss R} +
        \overline{E_{\sss L}} \,M_{\sss E}\, E_{\sss R} + h.c.
\label{L_Mass}
\end{equation}
where the mass matrices are related to the Yukawa coupling matrices
and Weinberg-Salam Higgs VEV by:
\begin{equation}
M = Y\; \frac{\sVEV{\phi_{\sss WS}}}{\sqrt{2}}
\label{mass-scale}
\end{equation}
We have chosen the normalisation from the Fermi coupling constant 
so that:
\begin{equation}
\sVEV{\phi_{\sss WS}} = 246\;\GeV\nn.
\label{WS-vev}
\end{equation}

In order to obtain the masses from the mass matrices, $M_{\sss U}$, 
$M_{\sss D}$ and $M_{\sss E}$, we must diagonalise them to find 
their eigenvalues. In particular we can find
unitary matrices, $V_{\sss U}$ for the up-type quarks, $V_{\sss D}$ 
for the down-type quarks and $V_{\sss E}$ for the charged leptons:
\begin{eqnarray}
V_{\sss U}^{\dagger}\,M_{\sss U}\,M_{\sss U}^{\dagger}\,V_{\sss U} & = &
\mbox{diag}\left( m_u^2, m_c^2, m_t^2 \right) \\
V_{\sss D}^{\dagger}\,M_{\sss D}\,M_{\sss D}^{\dagger}\,V_{\sss D} & = &
\mbox{diag}\left( m_d^2, m_s^2, m_b^2 \right) \\
\label{uni}
V_{\sss E}^{\dagger}\,M_{\sss E}\,M_{\sss E}^{\dagger}\,V_{\sss E} & = &
\mbox{diag}\left( m_e^2, m_\mu^2, m_\tau^2 \right)
\end{eqnarray}
The quark mixing matrix is then defined with these unitary matrices 
as~\cite{KM}: 
\begin{equation}
V_{{\rm\sss CKM}} = V_{\sss U}^{\dagger}\, V_{\sss D} \nn.
\end{equation}

\section{Model with many quantum numbers}
\indent

We have already investigated a model~\cite{NT1,NT2} which can 
predict not only quark and charged lepton quantities -- masses 
and mixing angles -- but also neutrino oscillations. This model 
has, as its back-bone, the property that there are generations 
(or families) not only for fermions but also 
for the gauge bosons, $\ie$ we have a generation (family) 
replicated gauge group namely 
\begin{equation}
  \label{eq:agut}
  \AGUT\nn,
\end{equation}
where $SMG$ denotes the Standard Model gauge group 
$\equiv SU(3)\times SU(2)\times U(1)$, $\times$ denotes the 
Cartesian product and $i$ runs through 
the generations. For the prediction of the charged particle masses and
mixings, the important part of the gauge group is the repetition of 
the Standard Model gauge group plus one extra $U(1)$ called $U(1)_f$, 
where $f$ denotes flavour~\cite{FLN, FNS, FNGS}. But for the extension 
to neutrino masses and mixings using the see-saw picture, it is 
necessary to introduce a right-handed neutrino and a gauged $B-L$ 
charge for each generation with the associated abelian gauge groups 
$U(1)_{\sss B-L,i}$ ($i=1,2,3$).
The just mentioned $U(1)_f$ abelian factor gets 
absorbed as a linear combination of the $B-L$ charge and the weak 
hypercharge abelian gauge groups for the different generations. 
Note that this family replicated gauge group, eq.~(\ref{eq:agut}), 
is the maximal gauge group under the following assumptions:
\begin{list}{\it\arabic{line})}{\usecounter{line}}
\item We only consider that part of the gauge group of Nature which 
acts non-trivially on the known 45 Weyl fermions of the Standard Model 
and the additional three heavy see-saw (right-handed) neutrinos. 
That is our gauge group is assumed to be a subgroup of $U(48)$.
\item We avoid any new gauge transformation that would transform a 
Weyl state from one irreducible representation of the Standard Model 
group into another irreducible representation: 
there is no gauge coupling unification.
\item  The gauge group does not contain any anomalies in the gauge 
symmetry -- neither gauge nor mixed anomalies. Note that otherwise 
the model becomes non-renormalisable.
\end{list}

\subsection{Gauge quantum numbers for the ``proto'' fermions at 
the fundamental scale}
\indent
In our model at the fundamental scale, which we take to be the 
Planck scale, there exist many bosons and fermions with practically
all quantum numbers we can ask for. But most of the fermions have vector
couplings, in the sense that they are described as Dirac particles 
from the Weyl point of view: they are combinations of left-handed and 
right-handed states with the same (gauge) quantum numbers. The left-over 
Weyl particles (in other word those without chiral partners) 
in our model are specified in more detail and are actually assumed to 
form a system of three proto-generations, each consisting of the 16 Weyl 
particles of  a usual Standard Model generation plus one see-saw particle.
In this way we can label these particle as proto-left-handed or
proto-right-handed $u$-quark, $d$-quark, electron $\etc$. To get the 
quantum numbers under our model gauge group for a given 
proto-irreducible representation, we proceed in the following way:
We note the generation number of the particle for which we want 
quantum numbers and we look up, in the Standard Model, what are the 
quantum numbers of the irreducible representation in question 
and what is the $B-L$ quantum number. For instance, if we want to find 
the quantum numbers of the proto-right-handed strange quark, we note 
that the quantum numbers of the right-handed strange quark in the 
Standard Model are weak hypercharge $y/2=-1/3$, 
singlet under $SU(2)$ and triplet under $SU(3)$, 
while $B-L$ is equal to the baryon number $=1/3$. Moreover, 
ignoring mixing angles, the 
generation is denoted as number $i = 2$.
The latter fact means that all the quantum numbers for $SMG_i$ 
$i=1,3$ are trivial. Also the baryon number minus lepton number for the
proto-generation number one and three are zero: only the quantum numbers 
associated with proto-generation two are non-trivial. Thus, in our model, 
the quantum numbers of the proto-right-handed strange quark are 
$y_2/2 = -1/3$, singlet under $SU(2)_2$, triplet under 
$SU(3)_2$ and $(B-L)_2=1/3$. For each proto-generation the following charge 
quantisation rule applies
\begin{equation}
  \label{eq:mod}
  \frac{t_i}{3}+\frac{d_i}{2} + \frac{y_i}{2} = 0~~{\rm (mod~1)}\nn,
\end{equation}
where $t_i$ and $d_i$ are the triality and duality for 
the $i$'th proto-generation gauge groups $SU(3)_i$ and $SU(2)_i$ 
respectively.

Combining eq.~(\ref{eq:mod}) with the principle of taking the smallest 
possible representation of the groups $SU(3)_i$ and $SU(2)_i$, it 
is sufficient to specify the six Abelian quantum numbers $y_i/2$ 
and $(B-L)_i$ in order to completely specify the gauge quantum 
numbers of the fields, $\ie$ of the Higgs fields and fermion fields.
Using this rule we easily specify the fermion representations as in 
Table $1$ (the representations of the Higgs fields will be 
given in subsection $5.1$, where we present the fermion 
mass matrices).

\begin{table}[!ht]
\caption{All $U(1)$ quantum charges for the proto-fermions in 
the model.}
\vspace{3mm}
\label{Table1}
\begin{center}
\begin{tabular}{|c||c|c|c|c|c|c|} \hline
& $SMG_1$& $SMG_2$ & $SMG_3$ & $U_{\sss B-L,1}$ & $U_{\sss B-L,2}$ & 
$U_{\sss B-L,3}$ \\ \hline\hline
$u_L,d_L$ &  $\frac{1}{6}$ & $0$ & $0$ & $\frac{1}{3}$ & $0$ & $0$ \\
$u_R$ &  $\frac{2}{3}$ & $0$ & $0$ & $\frac{1}{3}$ & $0$ & $0$ \\
$d_R$ & $-\frac{1}{3}$ & $0$ & $0$ & $\frac{1}{3}$ & $0$ & $0$ \\
$e_L, \nu_{e_{\sss L}}$ & $-\frac{1}{2}$ & $0$ & $0$ & $-1$ & $0$ & $0$ \\
$e_R$ & $-1$ & $0$ & $0$ & $-1$ & $0$ & $0$ \\
$\nu_{e_{\sss R}}$ &  $0$ & $0$ & $0$ & $-1$ & $0$ & $0$ \\ \hline
$c_L,s_L$ & $0$ & $\frac{1}{6}$ & $0$ & $0$ & $\frac{1}{3}$ & $0$ \\
$c_R$ &  $0$ & $\frac{2}{3}$ & $0$ & $0$ & $\frac{1}{3}$ & $0$ \\
$s_R$ & $0$ & $-\frac{1}{3}$ & $0$ & $0$ & $\frac{1}{3}$ & $0$\\
$\mu_L, \nu_{\mu_{\sss L}}$ & $0$ & $-\frac{1}{2}$ & $0$ & $0$ & $-1$ & 
$0$\\ $\mu_R$ & $0$ & $-1$ & $0$ & $0$  & $-1$ & $0$ \\
$\nu_{\mu_{\sss R}}$ &  $0$ & $0$ & $0$ & $0$ & $-1$ & $0$ \\ \hline
$t_L,b_L$ & $0$ & $0$ & $\frac{1}{6}$ & $0$ & $0$ & $\frac{1}{3}$ \\
$t_R$ &  $0$ & $0$ & $\frac{2}{3}$ & $0$ & $0$ & $\frac{1}{3}$ \\
$b_R$ & $0$ & $0$ & $-\frac{1}{3}$ & $0$ & $0$ & $\frac{1}{3}$\\
$\tau_L, \nu_{\tau_{\sss L}}$ & $0$ & $0$ & $-\frac{1}{2}$ & $0$ & $0$ & 
$-1$\\ $\tau_R$ & $0$ & $0$ & $-1$ & $0$ & $0$ & $-1$\\
$\nu_{\tau_{\sss R}}$ &  $0$ & $0$ & $0$ & $0$ & $0$ & $-1$ \\ 
\hline \hline
\end{tabular}
\end{center}
\end{table}

Note that each proto-generation gauge group $SMG_i \times U(1)_{B-L,i}$ 
is a subgroup of $SO(10)$, $\ie$ our gauge group eq.~(\ref{eq:agut}) is 
really a subgroup of $SO(10)^3$. That means the $i$'th proto-generation 
has its own subgroup of $SO(10)_i$. However, 
we do not take the gauge fields of these $SO(10)_i$ to exist, 
except for those corresponding to the
subgroups $SMG_i \times U(1)_{B-L,i}$.

\subsection{Breaking of the family replicated gauge group to the 
Standard Model}
\indent
The gauge group $\AGUT$ is at first spontaneously broken down at one 
or two orders of magnitude below the Planck scale, by 5 different 
Higgs fields, to the gauge group $SMG\times U(1)_{B-L}$ which is 
the diagonal subgroup of the original one:
\begin{equation}
  \label{eq:subagut}
  \left\{~(U,U,U)~|~U\subset SMG\times U(1)_{B-L}\right\} 
  \subseteq \AGUT\nn.
\end{equation}
We have to emphasize here that the gauge groups $SMG$ and 
$U(1)_{B-L}$ act similarly on all three families, $\ie$ they 
are not any more family replicated gauge groups but 
correspond to the usual gauge group of the Standard Model and the 
usual baryon number minus lepton number. This diagonal subgroup is 
further broken down by yet two more 
Higgs fields --- the Weinberg-Salam Higgs field $\Phi_{WS}$ and 
another Higgs field $\phi_{B-L}$ --- to 
$SU(3)\times U(1)_{em}$. The vacuum expectation value (VEV) of 
the $\phi_{B-L}$ Higgs field is taken to be about $10^{11}~\GeV$ 
and is designed to break the gauged $B-L$ quantum number. In 
other words the VEV $\sVEV{\phi_{B-L}}$ gives the see-saw scale.

Let us stress that we have {\em only one} Weinberg-Salam Higgs 
field $\Phi_{\sss WS}$, $\ie$ it only has one irreducible 
representation in our family replicated gauge group. 
We freely use both $\Phi_{\sss WS}$ 
and its Hermitian conjugate $\tilde\Phi_{\sss WS}$, 
which means that we have no supersymmetry in the model preventing  
one or the other from giving masses to the quarks and leptons.
Some of our predictions would be spoiled by introducing supersymmetry,  
because we need both a Higgs field and its Hermitian conjugate. 
With supersymmetry the number of Higgs fields and associated VEVs 
would have to be doubled;
in the special case of the Weinberg-Salam Higgs field, this means 
introducing the unknown parameter $\tan\beta$.

\subsubsection{Characterization of quark and charged lepton mass 
spectra}
\indent\ An important prediction of our model depends on the strongly 
non-supersymmetric feature of there being only one 
Weinberg-Salam Higgs field, but it is independent of the details of 
the other Higgs fields which break our gauge group down to the 
Standard Model. 
This predicted feature is that corresponding diagonal matrix elements 
in each of the three charged mass matrices, 
$M_{\sss U}, M_{\sss D}, M_{\sss E}$ 
and even for the Dirac neutrino mass matrix, $M_{\nu}^D$,
are order of magnitudewise the same~\cite{FLN,FNS}. 
%This feature related to the in a way remarkable feature of the 
%Standard Model that only one Weinberg-Salam Higgs field applies 
%to all mass matrices. 
The quantum number differences between the left- and 
right-handed Weyl fermions, between which a transition is needed to 
get these diagonal elements in our model, are $y_i/2 = \pm 1/2$ 
and $SU(2)_i$ representation equal to doublet, the rest being 
trivial, where $i$ is the proto-generation number of the diagonal 
element in question. 
Thus the quantum number violation needed, and therefore the order of 
magnitude resulting when all couplings are of order unity, will be 
the same for the diagonal element corresponding to a proto-family 
$i$ in each of the four left-to-right mass matrices, $M_{\sss U}$, 
$M_{\sss D}$, $M_{\sss E}$ and $M^D_{\nu}$ (the Dirac neutrino 
mass matrix).

The second and third family physical up-type  
quarks, $t$ and $c$, get their masses from two off-diagonal elements 
in $M_{\sss U}$ which dominate the diagonal ones in our model 
\cite{FLN,FNS}. 
So the above family degeneracy prediction then ends up becoming a 
prediction for the down-type quarks and leptons, simulating the  
simple $SU(5)$ GUT prediction ($m_b = m_{\tau}$, $m_s = m_{\mu}$, 
$m_d = m_e$), but {\em we only get it with respect to order 
of magnitude}. Thus our model can get the rough $SU(5)$ 
mass predictions, without having to suffer from the problem 
of needing, say, an extra $\underline{45}$ Higgs at the 
Weinberg-Salam Higgs scale and thereby varying the Clebsch-Gordon 
coefficients so as to cope with, what is honestly speaking, 
sheer disagreement for the simplest $SU(5)$ GUT. For the first 
family, in addition to the simulated GUT prediction, there is 
the degeneracy prediction that, when extrapolated to the Planck scale, 
$m_u\approx m_e$ order of magnitude-wise. This is an example of 
a prediction of our model that is sensitive to it \underline{not} 
being supersymmetric, because with supersymmety we would  
have two Weinberg-Salam Higgs fields and, with our philosophy 
that Higgs VEV's are likely to have their own order of 
magnitude, it would be difficult ever to get the prediction 
$m_u\approx m_e$.

Another regularity predicted from our model 
is the ``factorisation'' of the quark mixing angles 
\cite{FNS, FNGS}
\begin{equation}
  \label{eq:quarkmixing}
  V_{ub}\approx V_{us}\, V_{cb} \nn.
\end{equation}
This result mainly comes about because both $V_{ub}$ and 
$V_{us}$ contain, as a factor, similar Higgs field VEVs 
to take care of converting second family quantum numbers
into first family ones. Really five of the eleven predictions of 
our model are made up from these general rules: four from the 
family degeneracy predictions and one from the above 
factorisation.

%%%%%%%%%%%%%%%%%%%%%%%%%%%%%%%%%%%%%%%%%%%%%%%%%%%%%%%
\subsection{Introduction of Right-handed Majorana neutrinos}
\indent
In order to explain 
the neutrino oscillations, we have introduced three very heavy 
right-handed neutrinos into our model, which are mass-protected 
by the Higgs field, $\phi_{B-L}$, at an energy scale of 
about $10^{11}~\GeV$. We use the gauged $B-L$ charge to mass-protect
the right-handed neutrinos; in fact we use the total  -- 
diagonal -- one because we break $U(1)_{B-L, 1}\times U(1)_{B-L, 2}
\times U(1)_{B-L, 3}$ $\supset U(1)_{B-L}$ at a much higher 
energy scale, say about $10^{18}~\GeV$. Another new Higgs 
field, $\chi$, was also introduced in our previous see-saw 
model. This field plays the role of helping the VEV 
$\sVEV{\phi_{B-L}}$ to give non-zero effective mass terms 
for the see-saw neutrinos, by providing a transition between 
the right-handed tau-neutrino and the right-handed mu-neutrino. 
This transition coupling means that, with the new Higgs field 
$\chi$, we can obtain a large atmospheric neutrino
mixing angle.

However, unavoidably in the previous model, the solar mixing angle is 
in the region of the
small mixing angle MSW (SMA-MSW) 
solution, $\ie$ the solar mixing angle 
and the Cabibbo angle are characterised by the same parameter,
$\xi$, of order $1/10$. Furthermore the 
ratio of the solar neutrino mass squared difference to that for 
the atmospheric neutrino oscillations is given by $\xi^4$~\cite{NT1} 
without technical corrections~\cite{FF}. On the other hand, with these
technical corrections -- ``factorial factor corrections'' -- we
could manage to make a mass squared difference ratio consistent
with data for the SMA-MSW solutions~\cite{NT2}: 
due to the presence of a Higgs field 
$S$, whose VEV is of order one in Planck units, there are many 
choices of the quantum numbers of the other Higgs fields that only 
change the number of occurrences of this field $S$ in the fermion 
mass matrices. Moreover, one also has some 
freedom in the choice of the quantum numbers of the 
$\phi_{B-L}$ field, which spontaneously breaks the gauged $U(1)_{B-L}$ 
group and thereby gives the 
see-saw scale (about $10^{11}~\GeV$). In this way, we managed to get 
the mass squared difference ratio to be of zeroth order in $\xi$ 
and rather to be given by $S^8/4$, where the $S$ field VEV is 
close to unity.
However, the solar mixing angle could not be essentially changed, it 
remained of order $\xi$, and thus {\em only} fits the SMA-MSW region.

\subsection{Neutrino masses and mixing angles}
\label{sec:seesaw}
\indent\ The assumption of the existence of three right-handed Majorana 
neutrinos at a high scale\footnote{In the present model the right-handed 
neutrinos become massive by the action of the $\phi_{B-L}$ and 
$\chi$ Higgs fields together with another new Higgs field $\rho$. 
These massive right-handed neutrinos would all have decayed and be 
washed out completely by the present epoch in the evolution of the 
Universe.} gives rise to the addition of 
Majorana mass terms to the Lagrangian:
\begin{eqnarray}
  \label{lagrangian}
-\cL_{\sss\rm neutrino-mass} &\!=\!&
\bar{\nu}_L \,M^D_\nu \,\nu_R
+  \frac{1}{2}(\Bar{\nu_L})^{\sss c}\,M_{L}\, \nu_L
+  \frac{1}{2}(\Bar{\nu_R})^{\sss c}\,M_{R}\, \nu_R + h.c. \nonumber\\
&\!=\!&
\frac{1}{2} (\Bar{n_L})^{\sss c} \,M\, n_L + h.c.
\end{eqnarray}
where
\begin{equation}
n_L \!\equiv\! \left( \begin{array}{c}
    \nu_L \\
    (\nu_L)^{\sss c}
    \end{array} \right) \nn,\nn
M \!\equiv\! \left( \begin{array}{cc}
    M_L & M^D_\nu\\
    M^D_\nu & M_R
    \end{array} \right) \nn.
\end{equation}
\noindent
Here $M_\nu^D$ is the left-right transition mass term -- Dirac neutrino
mass term -- and  $M_L$ and $M_R$ are the isosinglet Majorana mass
terms of left-handed and right-handed neutrinos, respectively.  

Due to mass-protection by the Standard Model gauge symmetry, the 
left-handed Majorana mass terms, $M_L$, are negligible in our model 
with a fundamental scale set by the Planck mass~\cite{FNGS}. Then, 
naturally, the light neutrino mass matrix -- effective left-left 
transition Majorana mass matrix -- can be obtained via the see-saw 
mechanism~\cite{seesaw}:
\begin{equation}
  \label{eq:meff}
  M_{\rm eff} \! \approx \! M^D_\nu\,M_R^{-1}\,(M^D_\nu)^T\nn.
\end{equation}

In the framework of the three active neutrino model, the flavour
eigenstates $\nu_{\alpha}~(\alpha=e,\nu,\tau)$ are related to
the mass eigenstates $\nu_{i}~(i=1,2,3)$ in the vacuum by a 
unitary matrix $V_{\rm\sss MNS}$,
\begin{equation}
\label{eigen}
  \ket{\nu_{\alpha}} = \sum_i (V_{\rm\sss MNS})_{\alpha i} \ket{\nu_i}\nn.
\end{equation}
Here $V_{\rm\sss MNS}$ is the three-by-three 
Maki-Nakagawa-Sakata (MNS) mixing
matrix~\cite{MNS} which is parameterised by
\begin{eqnarray}
V_{\rm\sss MNS}&=& \left( \begin{array}{ccc}
  c_{13} c_{12}       & c_{13} s_{12}  & s_{13} e^{-i\delta_{13}} \\
- c_{23} s_{12} - s_{13} s_{23} c_{12} e^{i\delta_{13}}
& c_{23} c_{12} - s_{13} s_{23} s_{12} e^{i\delta_{13}}
& c_{13} s_{23} \\
    s_{23} s_{12} - s_{13} c_{23} c_{12} e^{i\delta_{13}}
& - s_{23} c_{12} - s_{13} c_{23} s_{12} e^{i\delta_{13}}
& c_{13} c_{23} \\
\end{array} \right)\nonumber\\
&&\nn\times\left( \begin{array}{ccc}
 e^{i\varphi} & 0 & 0 \\
 0 &  e^{i\psi} & 0\\
 0 &  0          & 1 \\
\end{array} \right) \nn,
\label{eq:MNS}
\end{eqnarray}
where $c_{ij} \equiv \cos\theta_{ij}$ and
$s_{ij} \equiv \sin\theta_{ij}$ and $\delta_{13}$ is a $CP$-violating
phase. Note that, due to the existence of Majorana neutrinos, we
have two additional $CP$-violating Majorana phases $\varphi$, 
$\psi$, which are also included in the MNS unitary mixing matrix.

In order to get predictions for the neutrino masses from the 
effective mass matrix, $M_{\rm eff}$, we have to diagonalise this 
matrix using a unitary matrix, $V_{\rm eff}$, to find the mass 
eigenvalues:
\begin{equation}
  \label{eq:mixingmatrix}
V_{\rm eff} M_{\rm eff} M_{\rm eff}^\dagger V_{\rm eff}^\dagger 
= {\rm diag}(m^2_1, m^2_2, m^2_3)\nn.
\end{equation}      
With the charged lepton unitary matrix $V_{\sss E}$, eq.~(\ref{uni}), 
we can then find the neutrino mixing matrix: 
\begin{equation}
  V_{\rm\sss MNS} = V_{\rm eff}^\dagger \,V_{\sss E}\nn.
\end{equation}%
Obviously, we should compare these theoretical predictions with
experimentally measured quantities, therefore we define:
\begin{eqnarray}
\label{eq:thtoex}
\Delta m^2_{\odot}&\equiv&m^2_2 - m^2_1\\
\Delta m^2_{\rm atm}&\equiv& m^2_3 - m^2_2\\
\tan^2\theta_{\odot} &\equiv& \tan^2\theta_{12} \\
\tan^2\theta_{\rm atm}&\equiv& \tan^2\theta_{23} \nn.      
\end{eqnarray}

Note that since we use the philosophy of order of 
magnitudewise predictions (see section~\ref{sec:FN}) with 
complex order one coupling constants, 
our model is capable of making predictions for these
three phases, the $CP$-violating phase $\delta_{13}$ 
and the two Majorana phases; put
simply, we assume all these phases are of order $\pi/2$, 
$\ie$ essentially maximal $CP$ violations.

%%%%%%%%%%%%%%%%%%%%%%%%%%%%%%%%%%%%%%%%%%%%%%%%%%%%%%%
\subsection{No-Go theorem for large mixing angle in previous model}
\indent
We have traced the reluctance of our previous see-saw models to fit 
the large mixing angle MSW solution to the following feature of 
the Dirac neutrino mass matrix, $M_\nu^D$: for every column, 
$\ie$ for all right-handed neutrinos in our notation, the first 
row elements -- left electron ones -- are smaller than the 
other matrix elements in the same column, by at least a 
factor of $\xi\approx 1/10$. With this
property, we can indeed prove that the solar mixing 
angle cannot be bigger than of order $\xi$, if we do not 
fine-tune the right-handed neutrino sector. 

Note that there is the possibility of getting a large solar neutrino 
mixing angle from the charged lepton sector, if it has big mixing 
relative to the proto-flavours~\cite{Fbled}. Our model, however, has 
an almost diagonal charged lepton mass matrix. Therefore we unavoidably 
obtain a small solar mixing angle, unless we re-arrange the Dirac 
neutrino mass matrix. That means that both the solar and atmospheric 
mixing angles must come from the Dirac neutrino sector in our present 
model.

Our no-go theorem states that, provided there is essentially no mixing 
in the charged lepton sector and that the Dirac neutrino mass matrix, 
$M_\nu^D$, obeys
\begin{equation}
  \label{eq:nogodirac}
\abs{(M_\nu^D)_{1 i}} \sleq  \abs{(M_\nu^D)_{2 i}} \xi 
\nn\nn {\rm and}\nn\nn
\abs{(M_\nu^D)_{1 i}} \sleq  \abs{(M_\nu^D)_{3 i}} \xi \nn\nn
{\rm for}~i=1,2,3\nn, 
\end{equation}
the solar mixing angle cannot be larger than of 
order $\xi$ ($\xi\approx 1/10$ in our previous model).

This no-go theorem is even harder to circumvent if one has an
$SO(10)$ gauge group, because the up-type mass matrix is then 
very strongly related with the Dirac neutrino mass matrix and also
the down-type mass matrix is similarly related with the charged 
lepton mass matrix~\cite{Ramond}. Really though it is necessary 
to exclude higher dimensional $SO(10)$ representations than say 
\underline{10} for Higgs fields at the Weinberg-Salam Higgs scale, 
in order to obtain the identity of the mass matrices~\cite{GJ}: 
\begin{equation}
  \label{eq:massre}
  M_{\sss U}=M^D_\nu\nn, \nn\nn M_{\sss D}=M_{\sss E}\nn.
\end{equation}
However, using this relationship, it is totally impossible to get a 
large solar mixing angle in the $SO(10)$ model.

\section{Discussion of modification}
\indent
In order to get an LMA-MSW solution to the solar neutrino problem, 
we have to re-arrange the Dirac neutrino sector~\cite{Isabella} as
we have already discussed in the previous section. 
We do this by the introduction of two new Higgs fields $\rho$ and 
$\omega$, which replace $S$ and $\xi$.

In our present model, we manage to make all the three elements of 
the first column (coupling to the first right-handed neutrino -- the 
lightest one in our case) in the Dirac neutrino mass matrix 
roughly equal in order of magnitude, $\ie$ the $(1,1)$, $(2,1)$ and 
even $(3,1)$ matrix elements are made the same order of magnitude.
The transition from the second to the first column corresponds to a 
shift in the generation $B-L$ quantum numbers, 
since it is given by the difference in the charges of the respective 
right-handed neutrinos. The new Higgs field, $\rho$, plays
this role; more precisely the third power of $\rho$ carries the 
quantum numbers required to make a 
transition form the first to the second column in the Dirac 
neutrino sector\footnote{The quantum numbers of this Higgs field 
can be found in Table~\ref{qc}.}. 
%On the place $(1,2)$ we get $(\omega/\rho)^3$.

In our new version of the model, the second and the third column 
in the Dirac neutrino mass matrix still obey the
condition of our ``no-go'' eq.~(\ref{eq:nogodirac}), which is actually 
very nice since it is needed to have a hope of getting a small CHOOZ 
angle $\theta_{13}$. If now all the elements in the first column would 
simply be obtained from the second by multiplication with 
$(\rho^{\dagger})^3$ as the quantum numbers at first suggest, 
this column would inherit the property of the first row element 
being small and we would not 
be able to get an LMA-MSW solution. However, we managed to get a need 
for the use of the $\rho$ field in the matrix element $(1,2)$ 
so that it has a factor of $\rho^3$ in it, and then in the transition 
to the first column we get rid of the $\rho^3$ factor rather 
than getting an extra factor of $(\rho^{\dagger})^3$. In this way 
we succeeded in making the ratio of matrix element 
$(1,1)$ to $(2,1)$ become bigger by a factor of 
$\rho^6$ than the ratio of $(1,2)$ to $(2,2)$. We 
really want the ratio of matrix element $(1,1)$ to $(2,1)$ to be of 
order unity, in order to obtain a large solar neutrino
mixing angle. This is arranged by introducing the Higgs field   
$\omega$ having a vacuum expectation value of about the same order 
of magnitude as $\rho$.

The value of the Cabibbo angle corresponded to the VEV of $\xi$ in 
the previous model. In the present model, it is given 
by the product $\omega\rho^\dagger$ whose VEV should thus be of 
order of $\xi\sim1/10$. From these considerations we can crudely 
estimate the VEVs of the new Higgs fields to be: 
$\omega\sim\rho\sim1/3$. 

\section{Method of numerical computation}
\label{sec:FN}
\indent
A very important assumption in our model is that, at the Planck scale,
we find a lot of different particles with many imaginable quantum 
numbers and having coupling constants which, when they are allowed, 
are complex numbers of order unity~\cite{FN}. This means that we 
assume essentially maximal $CP$ violation in all sectors, including 
the neutrino sector. Since we do not know the exact values of 
all these couplings we are, in general, only able to make predictions 
order of magnitudewise. According to this philosophy~\cite{FN}, 
we evaluate the product of mass-protecting Higgs VEVs required 
for each mass matrix element and provide it  
with a random complex number of order one as a factor. 
In this way, we simulate a long chain of fundamental Yukawa couplings 
and propagators making the transition corresponding to an 
effective Yukawa coupling in the Standard Model. In the numerical 
computation we then calculate the masses and mixing angles time 
after time, using different sets of random numbers and, in the 
end, we take the logarithmic average of the calculated quantities 
according to the following formula:
\begin{equation}
  \label{eq:avarage}
  \sVEV{m}=\exp\left(\sum_{i=1}^{N} \frac{\ln m_i}{N}\right) \nn. 
\end{equation}
Here $\sVEV{m}$ is what we take to be the prediction for one of the 
masses or mixing angles, $m_i$ is the result of the calculation
done with one set of random number combinations and $N$ is the total 
number of random number combinations used.

In order to find the best possible fit, we define 
a quantity which we call the goodness of fit (\mbox{\rm g.o.f.}). Since 
our model can only make predictions order of magnitudewise, this 
quantity \mbox{\rm g.o.f.} should only depend on the ratios of
the fitted masses (mass squared differences in the neutrino case) 
and mixing angles to the experimentally
determined masses and mixing angles:
\begin{equation}
\label{gof}
\mbox{\rm g.o.f.}\equiv\sum \left[\ln \left(
\frac{\sVEV{m}}{m_{\rm exp}} \right) \right]^2
\end{equation}
Here $\sVEV{m}$ are the fitted masses and mixing angles defined in 
eq.~(\ref{eq:avarage}) and $m_{\rm exp}$ are the
corresponding experimental values. The Yukawa coupling 
matrices are calculated at the fundamental scale,
which we take to be the
Planck scale. We use the first order renormalisation
group equations for the Standard Model to calculate 
the matrices at lower scales.
Running masses are calculated in terms of the Yukawa
couplings at $1~\GeV$ (see section \ref{sec:renom}).

\subsection{Quantum numbers of the Higgs fields}
\indent
The model we present in this article has exactly the same 
gauge group and gauge quantum numbers for the fermions 
as in earlier versions~\cite{NT1,NT2} of our see-saw model. 
It is only the system of Higgs fields which have different 
gauge quantum numbers and they are presented in Table~\ref{qc}.
The only essential change, even 
of the Higgs system, is that the fields $\omega$ and $\rho$ in 
the table replace the previous Higgs fields $S$ and $\xi$ and 
take on different quantum numbers. As can be
seen from Table~\ref{qc}, the fields $\omega$ and $\rho$ have only 
non-trivial quantum numbers with respect to the first
and second families. 
This choice of quantum numbers makes it possible to express a 
fermion mass matrix element involving the first family 
in terms of the corresponding element involving the second 
family, by the inclusion of an appropriate product of powers 
of $\rho$ and $\omega$. 

%%%%%%%%%%%%%%%%%%%%%%%%%%%%%%%%%%%%%%%%%%%%%%%%%%%%%%%
\begin{table}[!th]
\caption{All $U(1)$ quantum charges of the Higgs fields.}
\vspace{3mm}
\label{qc}
\begin{center}
\begin{tabular}{|c||c|c|c|c|c|c|} \hline
& $SMG_1$& $SMG_2$ & $SMG_3$ & $U_{\sss B-L,1}$ & $U_{\sss B-L,2}$ & $U_{\sss B-L,3}$ \\ \hline\hline   
$\omega$ & $\frac{1}{6}$ & $-\frac{1}{6}$ & $0$ & $0$ & $0$ & $0$\\
$\rho$ & $0$ & $0$ & $0$ & $-\frac{1}{3}$ & $\frac{1}{3}$ & $0$\\
$W$ & $0$ & $-\frac{1}{2}$ & $\frac{1}{2}$ & $0$ & $-\frac{1}{3}$ & $\frac{1}{3}$ \\
$T$ & $0$ & $-\frac{1}{6}$ & $\frac{1}{6}$ & $0$ & $0$ & $0$\\
$\chi$ & $0$ & $0$ & $0$ & $0$ & $-1$ & $1$ \\
$\phi_{\sss WS}$ & $0$ & $\frac{2}{3}$ & $-\frac{1}{6}$ & $0$ & $\frac{1}{3}$ & $-\frac{1}{3}$ \\
$\phi_{\sss B-L}$ & $0$ & $0$ & $0$ & $0$ & $0$ & $2$ \\
\hline
\end{tabular}
\end{center}
\end{table}   
%%%%%%%%%%%%%%%%%%%%%%%%%%%%%%%%%%%%%%%%%%%%%%%%%%%%%%%
In previous versions of the model, this role of the $\rho$ and $\omega$ 
fields was played by the fields $S$ and $\xi$ with the quantum 
number combinations (ordered as in Table 2):
\begin{eqnarray}
\label{eq:sandxi}
&S:&  \quad (\frac{1}{6},-\frac{1}{6},0,-1,-\frac{2}{3},\frac{2}{3})\\
&\xi:& \quad (\frac{1}{6},-\frac{1}{6},0,0,\frac{1}{3},-\frac{1}{3})
\end{eqnarray}
It is with these quantum numbers that one gets the ``no-go'' situation 
for the LMA-MSW solution, since the solar neutrino mixing angle then 
satisfies $\theta_{\odot}\sim\xi\sim V_{us}$. 
It turns out that, fitting with these 
``old'' quantum numbers, the vacuum expectation value of the field $S$ 
is close to being unity in fundamental (Planck) units. Once the 
Higgs field $S$ had a VEV of order unity, a large number of 
inessential modifications of the Higgs field quantum numbers became 
possible: one could add or subtract the quantum numbers of $S$ to/from 
any of the other proposed Higgs fields a large number of times,
without making any changes except in small details.  Therefore, in the 
previous work, it was necessary to consider and make fits using these 
other possibilities.

The new Higgs fields $\omega$ and $\rho$ turn out to have VEVs 
of the order of $1/3$. So, in the present model, there are no fields 
with a VEV of the order of unity and thus no such ambiguities in the 
choice of Higgs field quantum numbers.
In this way the ``new'' model escapes the ``discrete'' parameters of 
shuffling around the Higgs quantum numbers by multiples of those of 
$S$. So one now has a smaller amount of hidden 
fitting and a good fit should thus be considered a bit more impressive 
than in the previous model! The new model is in this way simplified 
compared to the old one. 

With the system of quantum numbers in 
Table~\ref{qc} one can easily evaluate, for a given mass matrix 
element, the numbers of Higgs field VEVs of the different types 
needed to perform the transition between the corresponding left- and 
right-handed Weyl fields. The results of calculating the products of 
Higgs fields needed, 
and thereby the order of magnitudes of the mass matrix elements 
in our model, are presented in the following mass matrices:

\noindent
the up-type quarks:
\begin{eqnarray}
M_{\sss U} \simeq \frac{\sVEV{(\phi_{\sss\rm WS})^\dagger}}{\sqrt{2}}
\hspace{-0.1cm}
\left(\!\begin{array}{ccc}
        (\omega^\dagger)^3 W^\dagger T^2
        & \omega \rho^\dagger W^\dagger T^2
        & \omega \rho^\dagger (W^\dagger)^2 T\\
        (\omega^\dagger)^4 \rho W^\dagger T^2
        &  W^\dagger T^2
        & (W^\dagger)^2 T\\
        (\omega^\dagger)^4 \rho
        & 1
        & W^\dagger T^\dagger
\end{array} \!\right)\label{M_U}
\end{eqnarray}  
\noindent
the down-type quarks:
\begin{eqnarray}
M_{\sss D} \simeq \frac{\sVEV{\phi_{\sss\rm WS}}}
{\sqrt{2}}\hspace{-0.1cm}
\left (\!\begin{array}{ccc}
        \omega^3 W (T^\dagger)^2
      & \omega \rho^\dagger W (T^\dagger)^2
      & \omega \rho^\dagger T^3 \\
        \omega^2 \rho W (T^\dagger)^2
      & W (T^\dagger)^2
      & T^3 \\
        \omega^2 \rho W^2 (T^\dagger)^4
      & W^2 (T^\dagger)^4
      & W T
                        \end{array} \!\right) \label{M_D}
\end{eqnarray}
\noindent %
the charged leptons:
\begin{eqnarray}        
M_{\sss E} \simeq \frac{\sVEV{\phi_{\sss\rm WS}}}
{\sqrt{2}}\hspace{-0.1cm}
\left(\hspace{-0.1 cm}\begin{array}{ccc}
    \omega^3 W (T^\dagger)^2
  & (\omega^\dagger)^3 \rho^3 W (T^\dagger)^2 
  & (\omega^\dagger)^3 \rho^3 W T^4 \chi \\
    \omega^6 (\rho^\dagger)^3  W (T^\dagger)^2 
  &   W (T^\dagger)^2 
  &  W T^4 \chi\\
    \omega^6 (\rho^\dagger)^3  (W^\dagger)^2 T^4 
  & (W^\dagger)^2 T^4
  & WT
\end{array} \hspace{-0.1cm}\right) \label{M_E}
\end{eqnarray}
\noindent
the Dirac neutrinos:
\begin{eqnarray}
M^D_\nu \simeq \frac{\sVEV{(\phi_{\sss\rm WS})^\dagger}}{\sqrt{2}}
\hspace{-0.1cm}
\left(\hspace{-0.1cm}\begin{array}{ccc}
        (\omega^\dagger)^3 W^\dagger T^2
        & (\omega^\dagger)^3 \rho^3 W^\dagger T^2
        & (\omega^\dagger)^3 \rho^3 W^\dagger  T^2 \chi\\
        (\rho^\dagger)^3 W^\dagger T^2
        &  W^\dagger T^2
        & W^\dagger T^2 \chi\\
        (\rho^\dagger)^3 W^\dagger T^\dagger \chi^\dagger
        &  W^\dagger T^\dagger \chi^\dagger
        & W^\dagger T^\dagger
\end{array} \hspace{-0.1 cm}\right)\label{Mdirac}
\end{eqnarray} 
\noindent %
and the Majorana (right-handed) neutrinos:
\begin{eqnarray}    
M_R \simeq \sVEV{\phi_{\sss\rm B-L}}\hspace{-0.1cm}
\left (\hspace{-0.1 cm}\begin{array}{ccc}
(\rho^\dagger)^6 (\chi^\dagger)^2
& (\rho^\dagger)^3 (\chi^\dagger)^2
& (\rho^\dagger)^3 \chi^\dagger \\
(\rho^\dagger)^3 (\chi^\dagger)^2
& (\chi^\dagger)^2 & \chi^\dagger \\
(\rho^\dagger)^3 \chi^\dagger & \chi^\dagger
& 1
\end{array} \hspace{-0.1 cm}\right ) \label{Mmajo}
\end{eqnarray}       
%23 dominant one.
%\begin{eqnarray}    
%M_R \simeq \sVEV{\phi_{\sss\rm B-L}}\hspace{-0.1cm}
%\left (\hspace{-0.1 cm}\begin{array}{ccc}
%(\rho^\dagger)^6 \chi^\dagger
%& (\rho^\dagger)^3 \chi^\dagger
%& (\rho^\dagger)^3  \\
%(\rho^\dagger)^3 \chi^\dagger
%& \chi^\dagger & 1 \\
%(\rho^\dagger)^3 & 1 & \chi
%\end{array} \hspace{-0.1 cm}\right )
%\end{eqnarray}       
In order to get the true model matrix elements, one must imagine 
that each matrix element is provided with an order of unity factor, 
which is unknown within our system of assumptions and which, as 
described above, is taken in our calculation as a complex random 
number, later to be logarithmically averaged over as in 
eq.~(\ref{eq:avarage}).

Note that the quantum numbers of our $6$ Higgs fields
are not totally independent. In fact there is a linear relation
between the quantum numbers of the three Higgs fields $W$, 
$T$ and $\chi$:
\begin{equation}
 \vec{Q}_\chi= 3\,\vec{Q}_W - 9\,\vec{Q}_T  
\end{equation}
where the 6 components of the charge vector $\vec{Q}$ correspond 
to the 6 columns of Table 2.
Thus the Higgs field combinations needed for a given transition are not
unique, and the largest contribution has to be selected for each matrix 
element in the above mass matrices. 

Furthermore, there is another remark: the symmetric
mass matrix -- for the Majorana neutrinos -- gives rise to the same 
off-diagonal term twice. The Feynman diagram for off-diagonal 
elements of the right-handed neutrino matrix is

\vspace*{.5cm}
%%%%%%%%%%%%%%%%%%%%%%%%%%%%%%%%%%%%%%%%%%%%%%%%%%%%%%%
\begin{equation}
  \label{eq:majlag}
\def\F#1#2{\fmfi{dots}{.8[vloc(__z),vloc(__#1)] %
-- .8[vloc(__z),vloc(__#2)]}}
2~(M_{\sss R})_{ij}~=  \quad \parbox{30mm}{
\unitlength=1mm
\begin{fmffile}{majofig1}
\begin{fmfgraph*}(30,20)
\fmfpen{thick}
\fmfleftn{i}{4}
\fmfrightn{j}{4}
\fmf{phantom}{i2,z,i3}
\fmf{dbl_plain_arrow}{i4,z,i1}
\fmf{plain}{j4,z,j3}
\fmf{plain}{j2,z,j1}                                                           
\fmffreeze
\fmfv{dec.shape=circle,dec.fill=shaded, dec.size=.2w}{z}
%label=$(M_{\sss R})_{ij}$,label.dist=-2.5mm}{z}
\fmfdraw
\F{j2}{j3}
\F{j3}{j4}
\fmfv{dec.shape=cross,dec.fill=1,dec.size=3thick}{j1,j2,j3,j4} 
\fmflabel{$\nu_{\! {\sss R}_i}$}{i4}
\fmflabel{$\nu_{\! {\sss R}_j}$}{i1}
\fmflabel{$\phi_{\sss B-L}$}{j1}
\fmflabel{$\rho^\dagger$}{j2}
\fmflabel{$\chi^\dagger$}{j4}
\fmflabel{$\rho^\dagger$}{j3}
\end{fmfgraph*}\end{fmffile}}
\quad \quad = \quad   
\parbox{30mm}{
\unitlength=1mm
\begin{fmffile}{majofig2}
\begin{fmfgraph*}(30,20)
\fmfpen{thick}
\fmfleftn{i}{4}
\fmfrightn{j}{4}
\fmf{phantom}{i2,z,i3}
\fmf{dbl_plain_arrow}{i4,z,i1}
\fmf{plain}{j4,z,j3}
\fmf{plain}{j2,z,j1}                                                           
\fmffreeze
\fmfv{dec.shape=circle,dec.fill=shaded, dec.size=.2w}{z}
%label=$(M_{\sss R})_{ji}$, label.dist=-2.5mm}{z}
\fmfdraw
\F{j2}{j3}
\F{j3}{j4}
\fmfv{dec.shape=cross,dec.fill=1,dec.size=3thick}{j1,j2,j3,j4} 
\fmflabel{$\nu_{\! {\sss R}_j}$}{i4}
\fmflabel{$\nu_{\! {\sss R}_i}$}{i1}
\fmflabel{$\phi_{\sss B-L}$}{j1}
\fmflabel{$\rho^\dagger$}{j2}
\fmflabel{$\chi^\dagger$}{j4}
\fmflabel{$\rho^\dagger$}{j3}
\end{fmfgraph*}\end{fmffile}}
\quad \quad=~2~(M_{\sss R})_{ji} \nn (i \not= j) \nn.
\end{equation}
\vspace*{.4cm}

Thus to avoid overcounting we just have to multiply off-diagonal 
elements of the right-handed Majorana mass matrix by a factor of 
$1/2$. However, in the Dirac mass matrix columns and rows are
related to completely different Weyl fields and, 
therefore, we do not need to worry about overcounting -- the 
off-diagonal elements should 
not be multiplied by an extra factor of $1/2$:

\vspace*{.4cm}
%%%%%%%%%%%%%%%%%%%%%%%%%%%%%%%%%%%%%%%%%%%%%%%%%%%%%%%
\begin{equation}
  \label{eq:dirlag}
\def\F#1#2{\fmfi{dots}{.8[vloc(__z),vloc(__#1)] %
-- .8[vloc(__z),vloc(__#2)]}}
(M^D)_{ij} ~= \quad  \quad \parbox{30mm}{
\unitlength=1mm
\begin{fmffile}{diracfig1}
\begin{fmfgraph*}(30,20)
\fmfpen{thick}
\fmfleftn{i}{4}
\fmfrightn{j}{4}
\fmf{phantom}{i2,z,i3}
\fmf{fermion}{i4,z,i1}
\fmf{plain}{j4,z,j3}
\fmf{plain}{j2,z,j1}                                                           
\fmffreeze
\fmfv{dec.shape=circle,dec.fill=shaded, dec.size=.2w}{z}
%label=$(M^D_{\nu})_{ij}$,label.dist=-2.5mm}{z}
\fmfdraw
\F{j2}{j3}
\F{j3}{j4}
\fmfv{dec.shape=cross,dec.fill=1,dec.size=3thick}{j1,j2,j3,j4} 
\fmflabel{$l_{\! {\sss L}_i}$}{i4} 
\fmflabel{$l_{\! {\sss R}_j}$}{i1}
\fmflabel{$\phi_{\sss WS}$}{j1}
\fmflabel{$W^\dagger$}{j2}
\fmflabel{$T$}{j4}
\fmflabel{$\rho$}{j3}
\end{fmfgraph*}\end{fmffile}}
\qquad \quad \not= \quad \quad
\parbox{30mm}{
\unitlength=1mm
\begin{fmffile}{diracfig2}
\begin{fmfgraph*}(30,20)
\fmfpen{thick}
\fmfleftn{i}{4}
\fmfrightn{j}{4}
\fmf{phantom}{i2,z,i3}
\fmf{fermion}{i4,z,i1}
\fmf{plain}{j4,z,j3}
\fmf{plain}{j2,z,j1}                                                           
\fmffreeze
\fmfv{dec.shape=circle,dec.fill=shaded, dec.size=.2w}{z}
%label=$(M^D_{\nu})_{ji}$, label.dist=-2.5mm}{z}
\fmfdraw
\F{j2}{j3}
\F{j3}{j4}
\fmfv{dec.shape=cross,dec.fill=1,dec.size=3thick}{j1,j2,j3,j4} 
\fmflabel{$l_{\! {\sss L}_j}$}{i4}
\fmflabel{$l_{\! {\sss R}_i}$}{i1}
\fmflabel{$\phi_{\sss WS}$}{j1}
\fmflabel{$W^\dagger$}{j2}
\fmflabel{$T$}{j4}
\fmflabel{$\rho$}{j3}
\end{fmfgraph*}\end{fmffile}}
\quad \quad =~(M^D)_{ji} \nn (i \not= j) \nn.
\end{equation}
%%%%%%%%%%%%%%%%%%%%%%%%%%%%%%%%%%%%%%%%%%%%%%%%%%%%%%%

\vspace*{.6cm}

The previous versions of our model predicted the Fritzsch 
relation~\cite{Fritzsch}
$V_{us}=\theta_{c}\approx\sqrt{m_d/m_s}$ (however only 
order of magnitudewise), provided that the VEV of the field $S$ was 
of order unity, $S\approx1$. With the above mass matrices, this relation 
is now replaced by a relation involving the solar neutrino mixing angle:
\begin{equation}
  \label{eq:newfritzsch}
  V_{us}=\theta_{c}\sim \left(\theta_{\odot}\right)^{-\frac{1}{3}}\, 
\left(\frac{m_d}{m_s}\right)^{\frac{2}{3}}.
\end{equation}

\subsection{Renormalisation group running of coupling constants}
\label{sec:renom}
\indent
It should be kept in mind that the effective Yukawa couplings for 
the Weinberg-Salam Higgs field, which 
are given by the Higgs field factors in the above mass matrices 
multiplied by some order unity factors (taken as random numbers), 
are the running (effective) Yukawa couplings at a scale {\em very 
close to the Planck scale}. Thus, in our calculations, we had to 
use the renormalisation group 
$\beta$-functions to run these couplings down to the experimentally 
observable scale, $\ie$ $\mu=1~\GeV$ where $\mu$ is the renormalisation 
point. This is because we 
took the charged fermion masses to be compared to ``measurements'' 
at the conventional scale of $1~\GeV$. In other words, what we 
take as input quark masses are the current algebra masses, 
corresponding to running masses at $1~\GeV$, except for the 
top quark. We used the top quark pole mass instead:
\begin{equation}
M_t = m_t(M)\left(1+\frac{4}{3}\frac{\alpha_s(M)}{\pi}\right)\nn,
\end{equation}
where we set $M=180~\GeV$ as an input, for simplicity.

Using the notation in eq.~(\ref{L_Higgs}), we can define the one-loop 
$\beta$ functions for the gauge couplings and the charged fermion 
Yukawa matrices~\cite{pierre} as follows:
\begin{eqnarray}
  \label{eq:recha}
16 \pi^2 {d g_{1}\over d  t} &\!=\!& \frac{41}{10} \, g_1^3 \nonumber\\
16 \pi^2 {d g_{2}\over d  t} &\!=\!& - \frac{19}{16} \, g_2^3  \nonumber\\
16 \pi^2 {d g_{3}\over d  t} &\!=\!& - 7 \, g_3^3  \nonumber\\
16 \pi^2 {d Y_{\sss U}\over d  t} &\!=\!& \frac{3}{2}\, 
\left( Y_{\sss U} (Y_{\sss U})^\dagger
-  Y_{\sss D} (Y_{\sss D})^\dagger\right)\, Y_{\sss U} 
+ \left\{\, Y_{\sss S} - \left(\frac{17}{20} g_1^2 
+ \frac{9}{4} g_2^2 + 8 g_3^2 \right) \right\}\, Y_{\sss U}\\
16 \pi^2 {d Y_{\sss D}\over d  t} &\!=\!& \frac{3}{2}\, 
\left( Y_{\sss D} (Y_{\sss D})^\dagger
-  Y_{\sss U} (Y_{\sss U})^\dagger\right)\,Y_{\sss D} 
+ \left\{\, Y_{\sss S} - \left(\frac{1}{4} g_1^2 
+ \frac{9}{4} g_2^2 + 8 g_3^2 \right) \right\}\, Y_{\sss D} \nonumber\\
16 \pi^2 {d Y_{\sss E}\over d  t} &\!=\!& \frac{3}{2}\, 
\left( Y_{\sss E} (Y_{\sss E})^\dagger \right)\,Y_{\sss E} 
+ \left\{\, Y_{\sss S} - \left(\frac{9}{4} g_1^2 
+ \frac{9}{4} g_2^2 \right) \right\}\, Y_{\sss E}\nonumber\\
Y_{\sss S} &\!=\!& {\Tr}(\, 3\, Y_{\sss U}^\dagger\, Y_{\sss U} 
+  3\, Y_{\sss D}^\dagger \,Y_{\sss D} +  Y_{\sss E}^\dagger\, 
Y_{\sss E}\,)  \nonumber\nn,
\end{eqnarray}
where $t=\ln\mu$.

In order to run the the renormalisation group
equations down to $1~\GeV$, we use the following initial values:
\begin{eqnarray}
U(1):   \quad & g_1(M_Z) = 0.462\nn, \quad & g_1(M_{\rm Planck}) = 0.614\\
SU(2):  \quad & g_2(M_Z) = 0.651\nn, \quad & g_2(M_{\rm Planck}) = 0.504\\
SU(3):  \quad & g_3(M_Z) = 1.22 \nn, \quad & g_3(M_{\rm Planck}) = 0.491
\end{eqnarray}

\subsection{The renormalisation group equations for the effective 
neutrino mass matrix}
\indent

The effective light neutrino masses are given by an irrelevant, 
non-renormalisable dimension $5$ term~\cite{BC,ADKLRCW}:
\begin{eqnarray}
\Delta L_{\rm eff} = {1\over2}~C^{ij} \; (\epsilon_{ab} H_a l_b^i )
\, (\epsilon_{cd} H_c l_d^j) \nn,
\end{eqnarray}
where $l_a^i$ are left-handed Weyl lepton fields with the flavour
index $i$ and $SU(2)$ weak isospin index $a$, and $C^{ij}$ is 
a symmetric matrix of coefficients:
\begin{eqnarray}
C^{ij}(\mu) = Y_\nu^{ki} (M_{R}^{-1})^{kl} Y_\nu^{lj}\nn.
\end{eqnarray}
Here $M_R$ is the Majorana mass matrix of the right-handed Majorana 
neutrinos, and $Y_\nu$ is the Dirac neutrino Yukawa coupling matrix.

The renormalisation group equations for the symmetric matrix, 
$C^{ij}$, are given by
\begin{equation}
16 \pi^2 {dC^{ij}\over d  t}
= ( - 3 g_2^2 + 2 \lambda + 2 Y_{\sss S} ) \; C^{ij}
- {3\over 2} \left( C^{ik}(Y_{\sss E}^\dagger)^{kl} (Y_{\sss E})^{lj} 
+ (Y_{\sss E}^\dagger)^{lk} (Y_{\sss E})^{ki}\; C^{lj} \right)\nn,
\end{equation}
where $\lambda$ is the Weinberg-Salam Higgs self-coupling constant and 
\begin{equation}
Y_{\sss S}={\Tr}( \,3\,Y_{\sss U}^\dagger \,Y_{\sss U} 
+ 3 \,Y_{\sss D}^\dagger \,Y_{\sss D} + Y_{\sss E}^\dagger \,
Y_{\sss E}\,)  \nn.
\end{equation}
The mass of the Standard Model Higgs boson is given by 
$M_H^2 = \lambda \sVEV{\phi_{WS}}^2$ and, for definiteness, we 
take $M_H = 115~\GeV$ thereby fixing the value 
of the Higgs self-coupling $\lambda=0.2185$.
These evolution equations can be rewritten using the elements of the 
light neutrino effective mass matrix $M_{\rm eff}$ as running 
quantities:
\begin{equation}
\label{eq:remeff}
16 \pi^2 {d M_{\rm eff} \over d  t}
= ( - 3 g_2^2 + 2 \lambda + 2 Y_{\sss S} ) \,M_{\rm eff}
- {3\over 2} \left( M_{\rm eff}\, ( Y_{\sss E} Y_{\sss E}^\dagger )^T 
+ ( Y_{\sss E} Y_{\sss E}^\dagger ) \,M_{\rm eff}\right) \nn.
\end{equation}
Note that the renormalisation group equations are used to evolve 
the effective neutrino mass matrix from the see-saw 
sale, set by $\sVEV{\phi_{B-L}}$ in our model, to $1~\GeV$. 
%$\ie$, the energy scale is taken to be the interval 
%$\mu\in\left[\sVEV{\phi_{B-L}}, 1~\GeV\right]$ in neutrino sector. 
We should emphasize that we have used the approximation of 
ignoring the running of the Dirac neutrino Yukawa coupling constants 
between the Planck scale and the see-saw scale; however, this 
effect is small and so this approximation should be good enough for 
our order of magnitude calculations.

\section{Numerical results}
\indent
Using the three charged quark-lepton mass matrices and the effective 
neutrino mass matrix together with the renormalisation group 
equations, eqs.~(\ref{eq:recha}) and (\ref{eq:remeff}), we made 
a fit to all the fermion quantities in Table 3 varying just 6 Higgs 
fields VEVs. We averaged over $N=10,000$ complex order unity random 
number combinations (see eq.~\ref{eq:avarage}). These complex numbers
are chosen to be the exponential of a number picked from a Gaussian 
distribution, with mean value zero and standard deviation one, 
multiplied by a random phase factor. We varied the 6 free 
parameters and found the best fit, corresponding to the lowest value 
for the quantity \mbox{\rm g.o.f.} defined in eq.~(\ref{gof}), with the 
following values for the VEVs:
\begin{eqnarray} 
\label{eq:VEVS} 
&&\sVEV{\phi_{\sss WS}}= 246~\GeV\nn,  
\nn\sVEV{\phi_{\sss B-L}}=1.64\times10^{11}~\GeV\nn, 
\nn\sVEV{\omega}=0.233\nn,\nonumber\\
&&\nn\sVEV{\rho}=0.246\nn,\nn\sVEV{W}=0.134\nn,
\nn\sVEV{T}=0.0758\nn,\nn\sVEV{\chi}=0.0737\nn,
\end{eqnarray}
where, except for the Weinberg-Salam Higgs field and 
$\sVEV{\phi_{\sss B-L}}$, the VEVs are expressed in Planck units. 
Hereby we have considered that the Weinberg-Salam Higgs field VEV is 
already fixed by the Fermi constant.
The results of the best fit, with the VEVs in eq.~(\ref{eq:VEVS}), 
are shown in Table~\ref{convbestfit} and the fit has  
$\mbox{\rm g.o.f.}=3.63$. 

We have $11=17 - 6$ degrees of freedom -- predictions -- leaving each of 
them with a logarithmic error of
$\sqrt{3.63/11}\simeq0.57$, which is very close to the 
theoretically  expected value $0.64$~\cite{FF}. This means, in other words, 
that we can fit {\rm all quantities} within a factor 
$1.78\simeq\exp\left(\sqrt{3.63/11}\right)$ of the experimental value.
However, we do not count the $42$ complex order unity random 
numbers entering the mass matrices eqs.~$(32)$-$(36)$ as parameters 
when discussing predictions, since we {\em do not adjust them}. We only
use them as a \underline{calculational technique} to
avoid ``unnatural'' matrices which would be degenerate 
in our calculations, if we did not have random number coefficients.

%%%%%%%%%%%%%%%%%%%%%%%%%%%%%%%%%%%%%%%%%%%%%
\begin{table}[!t]
\caption{Best fit to conventional experimental data.
All masses are running
masses at $1~\GeV$ except the top quark mass which is the pole mass.
Note that we use the square roots of the neutrino data in this 
Table, as the fitted neutrino mass and mixing parameters 
$\sVEV{m}$, in our goodness of fit ($\mbox{\rm g.o.f.}$) definition, 
eq.~(\ref{gof}).}
\begin{displaymath}
\begin{array}{|c|c|c|}
\hline\hline
 & {\rm Fitted} & {\rm Experimental} \\ \hline
m_u & 4.4~\MeV & 4~\MeV \\
m_d & 4.3~\MeV & 9~\MeV \\
m_e & 1.0~\MeV & 0.5~\MeV \\
m_c & 0.63~\GeV & 1.4~\GeV \\
m_s & 340~\MeV & 200~\MeV \\
m_{\mu} & 80~\MeV & 105~\MeV \\
M_t & 208~\GeV & 180~\GeV \\
m_b & 7.2~\GeV & 6.3~\GeV \\
m_{\tau} & 1.1~\GeV & 1.78~\GeV \\
V_{us} & 0.093 & 0.22 \\
V_{cb} & 0.027 & 0.041 \\
V_{ub} & 0.0025 & 0.0035 \\ \hline
\Delta m^2_{\odot} & 9.5 \times 10^{-5}~\eV^2 &  4.5 \times 10^{-5}~\eV^2 \\
\Delta m^2_{\rm atm} & 2.6 \times 10^{-3}~\eV^2 &  3.0 \times 10^{-3}~\eV^2\\
\tan^2\theta_{\odot} &0.23 & 0.35\\
\tan^2\theta_{\rm atm}& 0.65 & 1.0\\
\tan^2\theta_{13}  & 4.8 \times 10^{-2} & \sleq~2.6 \times 10^{-2}\\
\hline\hline
\mbox{\rm g.o.f.} &  3.63 & - \\
\hline\hline
\end{array}
\end{displaymath}
\label{convbestfit}
\end{table}
%%%%%%%%%%%%%%%%%%%%%%%%%

Unlike in older versions of the model, the first and second 
family sub-matrix of $M_{\sss D}$ is now dominantly diagonal. In 
previous versions of the model this submatrix satisfied the 
order of magnitude factorisation condition $(M_{\sss D})_{12}\cdot
(M_{\sss D})_{21}$ $\approx$ $(M_{\sss D})_{11}\cdot(M_{\sss D})_{22}$; 
thus the down quark mass $m_d$ received two contributions 
(off-diagonal as well as diagonal) of the same order of magnitude 
as the up quark mass $m_u$. This extra off-diagonal contribution to 
$m_d$ of course improved the goodness of the fit to the masses of 
the first family, since phenomenologically $m_d \approx 2~m_u$. 
However, in the present version of the model with the $\omega$ 
and $\rho$ Higgs fields, the off-diagonal element
$(M_{\sss D})_{21}$ becomes smaller and we are left with a full 
order of magnitude degeneracy of the first family masses, even 
including the down quark. 
Furthermore, our expectation from section 4 that $\sVEV{\omega} \sim
\sVEV{\rho} \sim1/3$ tends to overestimate the first family masses. 
So the result of our fit, eq.~(\ref{eq:VEVS}), is to take $\sVEV{W}$ 
somewhat smaller than in our previous models and $\sVEV{\omega} \sim
\sVEV{\rho} < 1/3$. 
Consequently our best fit values for the charm quark mass $m_c$ 
and the Cabibbo angle $V_{us}$ are smaller than in our 
previous fits to the charged fermion masses \cite{FNS,FNGS,FF}. 
Nonetheless, as mentioned above, our present best fit agrees 
with the experimental data within the theoretically expected 
uncertainty of about $64\%$ and is, therefore, as good as can be 
expected from an order of magnitude fit.  

Experimental results on the values of neutrino mixing angles 
are often presented in terms of the function $\sin^22\theta$ 
rather than $\tan^2\theta$ (which, contrary to $\sin^22\theta$, 
does not have a maximum at $\theta=\pi/4$ and thus still varies 
in this region).
Transforming from $\tan^2\theta$ variables to $\sin^22\theta$ 
variables, our predictions for the neutrino mixing angles become:
\begin{eqnarray}
  \label{eq:sintan}
 \sin^22\theta_{\odot} &\!=\!&0.61\nn,\\
 \sin^22\theta_{\rm atm} &\!=\!& 0.96\nn, \\
 \sin^22\theta_{13} &\!=\!& 0.17\nn.
\end{eqnarray}  
We also give here our predicted hierarchical neutrino mass 
spectrum:
\begin{eqnarray}
m_1 &\!=\!&  4.9\times10^{-4}~~\eV\nn, 
\label{eq:neutrinomass1}\\
m_2 &\!=\!&  9.7\times10^{-3}~~\eV\nn, 
\label{eq:neutrinomass2}\\
m_3 &\!=\!&  5.2\times10^{-2}~~\eV\nn.
\label{eq:neutrinomass3} 
\end{eqnarray}

Compared to the experimental data these predictions are 
excellent: all of our order of magnitude neutrino predictions lie 
inside the $99\%$ C.L. border determined from phenomenological fits 
to the neutrino data, even including the CHOOZ upper bound.
On the other hand, our prediction of the solar mass squared 
difference is about a factor of $2$ larger than the global fit
data even though the prediction is inside of the LMA-MSW region, 
giving a contribution to our goodness of fit of \mbox{\rm g.o.f.} 
$\approx 0.14$. Our CHOOZ angle also turns out to be 
about a factor of $2$ larger than the experimental limit at 
$90\%$ C.L., corresponding to another contribution of \mbox{\rm g.o.f.} 
$\approx 0.14$. In summary our predictions for the neutrino sector 
agree extremely well with the data, giving a contribution of only 
0.34 to \mbox{\rm g.o.f.} while the charged fermion sector contributes 
3.29 to \mbox{\rm g.o.f.}.

\subsection{CHOOZ angle and three flavour analysis}
\label{sec:chooz}
\indent\ The combination of
the results from atmospheric neutrino experiments~\cite{SK}
and the CHOOZ reactor experiment~\cite{CHOOZ} constrains 
the first- and third-generation mixing angle to be small,
$\ie$ the $3 \sigma$ upper bound is given by 
$\tan^2\theta_{13}~\sleq~0.06$.
This limit was obtained from a three flavour neutrino analysis 
(in the five dimensional parameter space -- 
$\theta_\odot$, $\theta_{13}$, $\theta_{\rm atm}$,
$\Delta m^2_{\odot}$ and $\Delta m^2_{\rm atm}$), 
using all the solar and atmospheric neutrino data and
based on the assumption that
neutrino masses have a hierarchical structure, $\ie$ 
$\Delta m^2_{\odot}\ll\Delta m^2_{\rm atm}$~\cite{cc3}.

However, the solar neutrino data in Table 3 come from a global 
two flavour analysis, which means that the first- and 
third-generation mixing angle is essentially put 
equal to zero, $\ie$ the dependence of $\theta_{13}$
on the solar neutrino parameters have been ignored. In 
principle we should, of course, fit to neutrino parameters from a 
three flavour analysis.
Recently, global three flavour analyses were 
performed~\cite{fogli2,cc2,serguey1} and they showed a significant 
influence of the non-zero CHOOZ angle on the solar neutrino
mass squared difference and mixing angle\footnote{In~\cite{serguey1} 
the relatively large solar neutrino mass squared difference lying 
in the LMA-MSW region (with the condition 
$\Delta m^2_{\odot}\sgeq10^{-4}~\eV^2$), 
the solar mixing angle and the CHOOZ reactor experiment data
were analysed using the three flavour analysis method.} and vice 
versa: if the CHOOZ angle
becomes far from zero then the solar mixing angle becomes
smaller. This effect is more significant for the larger 
$\Delta m^2_{\odot}$ values. Because of this correlation,
our fit to the neutrino data is even somewhat better than 
that suggested by the \mbox{\rm g.o.f.} value; even including 
the CHOOZ angle our neutrino fit is extremely good.

\subsection{$CP$ violation}
\indent\ We have fitted all the fermion masses and their mixing 
angles and therefore have predictions for the CKM
and MNS mixing matrices, in the quark sector and in the lepton 
sector respectively, including $CP$ violating phases of order 
unity. In this subsection, we will first consider the size of 
$CP$ violation in the quark sector and then the electron 
``effective Majorana mass'' responsible for neutrinoless 
double beta decay.

The Jarlskog invariant $J_{\sss CP}$ provides a measure of the 
amount of $CP$ violation in the quark sector~\cite{cecilia} and, 
in the approximation of setting cosines of mixing angles to unity, 
is just twice the area of the unitarity triangle:
\begin{equation}
  \label{eq:jarkskog}
  J_{\sss CP}=V_{us}\,V_{cb}\,V_{ub}\,\sin \delta \nn,
\end{equation}
where $\delta$ is the $CP$ violation phase in the CKM matrix.
In our model the quark mass matrix elements have random phases, 
so we expect $\delta$ (and also the three angles $\alpha$, 
$\beta$ and $\gamma$ of the unitarity triangle) to be of 
order unity and, taking an average value of 
$|\sin\delta| \approx 1/2$, the area of the 
unitarity triangle becomes
\begin{equation}
  \label{eq:jarkskog*0.5}
  J_{\sss CP}\approx \frac{1}{2}\,V_{us}\,V_{cb}\,V_{ub}\nn.
\end{equation}
Using the best fit values for the CKM elements from 
Table~\ref{convbestfit}, we predict 
$J_{\sss CP} \approx 3.1\times10^{-6}$ to be compared with 
the experimental value $(2-3.5)\times10^{-5}$. 
Since our result for the Jarlskog invariant  
is the product of four quantities, we do not expect the 
usual $\pm64\%$ logarithmic uncertainty but rather 
$\pm\sqrt{4}\cdot64\%=128\%$ logarithmic
uncertainty. This means our result deviates from the 
experimental value by 
$\log \frac{2.7 \times 10^{-5}}{3.1 \times 10^{-6}}/1.28$ 
= 1.7 ``standard deviations''. 

Another prediction, which can also be made from this model, is 
the electron ``effective Majorana mass'' -- the parameter in  
neutrinoless beta decay -- defined by: 
\begin{equation}
\label{eq:mmajeff}
\abs{\sVEV{m}} \equiv \abs{\sum_{i=1}^{3} U_{e i}^2 \, m_i} \nn,
\end{equation}
where $m_i$ are the masses of the neutrinos $\nu_i$ 
and $U_{e i}$ are the MNS mixing matrix elements for the 
electron flavour to the mass eigenstates $i$. We can 
substitute values for the neutrino masses $m_i$ from 
eqs.~(\ref{eq:neutrinomass1}-\ref{eq:neutrinomass3}) and for the 
fitted neutrino mixing angles from Table 3 into the left hand side of 
eq.~(\ref{eq:mmajeff}). 
As already mentioned, the $CP$ violating phases in the MNS mixing 
matrix are essentially random in our model. So we combine the 
three terms in eq.~(\ref{eq:mmajeff}) by taking the square root of 
the sum of the modulus squared of each term, which gives
our prediction:
\begin{equation}
  \label{eq:meffresult}
  \abs{\sVEV{m}} \approx 3.1\times 10^{-3}~~\eV\nn.
\end{equation}

Although the Jarlskog invariant and the effective Majorana electron 
neutrino mass have been calculated from the best fit parameters in 
Table 3, it is also possible to calculate them directly while 
making the fit. So we have calculated $J_{\sss CP}$ and $\abs{\sVEV{m}}$ 
for $N=10,000$ complex order unity random 
number combinations. Then we took the logarithmic average 
of these $10,000$ samples of $J_{\sss CP}$ and $\abs{\sVEV{m}}$ 
and obtained the following results:
\begin{eqnarray}
  \label{eq:jcpabsm}
   J_{\sss CP}&=& 3.1\times 10^{-6} \nn,\\
    \abs{\sVEV{m}}&=& 4.4\times 10^{-3}~~\eV\nn.
    \label{eq:mmaj2} 
\end{eqnarray}  
in good agreement with the values given above.

We should mention here that our effective Majorana mass parameter 
eq.~(\ref{eq:mmaj2}), of course,
respects the upper limit presented in ref.~\cite{serguey2}.

\section{Conclusion}
\indent\ We have developed an older version of our model, with the 
purpose of making it fit the experimentally favored LMA-MSW 
solution rather than the SMA-MSW solution for solar neutrino 
oscillations. In the older version, the magnitudes of the
solar mixing angle $\theta_{\odot}$ and the Cabibbo angle 
$\theta_c$ are both characterised by the VEV of the Higgs field 
$\xi \sim 1/10$ and thus the previous model could only be made 
compatible with the SMA-MSW solution.
The required modification of the model was achieved 
by replacing the fields $S$ and $\xi$ in the previous model by another 
pair of Higgs fields: $\omega$ and $\rho$ having non-trivial and 
opposite quantum numbers with 
respect to the family one and family two gauge groups, while having 
trivial family three gauge quantum numbers. In this way an 
excellent fit to the LMA-MSW solution is obtained. The price 
paid for the greatly improved neutrino mass matrix fit -- the 
neutrino parameters now contribute only very little to the 
\mbox{\rm g.o.f.} -- is a slight deterioration in the fit to the 
charged fermion 
mass matrices. In particular the predicted values of the quark masses 
$m_d$ and $m_c$ and the Cabibbo angle $V_{us}$ are reduced compared to 
our previous fits. However the overall fit agrees with the seventeen 
measured quark-lepton mass and mixing angle parameters in Table 3 
within the theoretically expected uncertainty~\cite{FF} of about 
$64\%$; it is a perfect fit order of magnitudewise. 

It should be remarked that our model provides 
an order of magnitude fit{}/{}understanding of all the effective
Yukawa couplings of the Standard Model and the neutrino 
oscillation parameters in terms of only $6$ parameters -- the Higgs 
field vacuum expectation values. So we can say that we fit all 
the parameters of the Standard Model and neutrino oscillations, 
except for the gauge coupling constants and the Higgs mass and 
its self-coupling. Actually we should note here that even the 
gauge coupling  constants may be derived from order one quantities
in the following sense: We postulate that the gauge couplings
-- and here, perhaps a bit arbitrarily at first, let us say that we 
take these to be $g_1$, $g_2$ and $g_3$ -- are of order unity at the 
Planck scale. That means that the corresponding inverse $\alpha$'s 
(the inverse fine structure constants) are of order $4\pi=12.5$. Now, 
according to our model, the 
Standard Model gauge groups which we see 
experimentally are the diagonal subgroups of a cross product of three 
replicas of the same mathematical group; one cross product factor for
each family. The formula for calculating the inverse $\alpha$
for the diagonal subgroup, $\alpha_{DS}$, is:
\begin{equation}
\alpha_{DS}^{-1}=\sum_{i=1}^3 {\alpha_i}^{-1}\nn. 
\end{equation} 
Thus the order unity assumption leads in our model to a Standard Model 
inverse $\alpha$ at the Planck scale 
being of the order of $3\cdot12.5=37.5$. This agrees very well 
with the experimentally measured Standard Model couplings, when 
extrapolated to the Planck scale using the renormalisation group 
equations of section 5.2, being respectively:
$\alpha_1^{-1}=55.5$, $\alpha_2^{-1}=49$, $\alpha_3^{-1}=54$. 
This observation can be further elaborated and 
converted into an exact prediction~\cite{MPP}, by using the 
so-called Multiple Point Principle (MPP). 
The phase transition couplings used in this MPP can be taken as 
the definition of couplings being of order unity and, with this 
choice, it turns out that it is indeed the $g_1$, $g_2$ and $g_3$
that are really of order unity rather than, say, the $\alpha_1$, 
$\alpha_2$ and $\alpha_3$.

In a similar way the application~\cite{HiggsFNT} of the MPP, requiring 
degenerate minima in the Weinberg-Salam Higgs effective 
potential, can lead to a Higgs field 
self-coupling which is of a similar order to the squares of the $g_i$'s, 
$\ie$ $\lambda\approx g_i^2$.

Thus, including this sort of argument, our model gives an order of 
magnitude understanding of all the Standard Model gauge, Higgs and 
Yukawa couplings, and even also of the beyond the  
Standard Model neutrino oscillation parameters.

%%%%%%%%%%%%%%%%%%%%%%%%%%%%%%%%%%%%%%%%%%%%%%%%%%%%%%%
\section*{Acknowledgements}

We wish to thank M.~C.~Gonzalez-Garcia, S.~T.~Petcov, 
C.~Pe{\~n}a-Garay, and P.~Ramond for useful discussions. 

C.D.F. and H.B.N. thank the EU commission for grants 
NTAS-RFBR-95-0567 and INTAS 93-3316(ext).
H.B.N. also wishes to thank the EU commission for grants 
SCI-0430-C (TSTS) and CHRX-CT-94-0621. C.D.F. thanks PPARC 
for a travel grant to attend a Bled workshop in July 2001. 
Y.T. thanks the Frederikke L{\o}rup f{\o}dt Helms Mindelegat for 
a travel grant to attend the EPS HEP 2001 and the 4th Bled workshop.

%\newpage 


\begin{thebibliography}{99}
%
\bibitem{SNO} Q.~R.~Ahmad {\it et al.}, SNO Collaboration, 
Phys.\ Rev.\ Lett.\  {\bf 87} (2001) 071301.
%%CITATION = NUCL-EX 0106015;%%
%
\bibitem{chlorine} B.~T.~Cleveland {\it et al.}, 
Astrophys.\ J.\  {\bf 496} (1998) 505.
%%CITATION = ASJOA,496,505;%%
%
\bibitem{sage}  J.~N.~Abdurashitov {\it et al.}, SAGE Collaboration,
Phys.\ Rev.\ C {\bf 60} (1999) 055801.
%%CITATION = ASTRO-PH 9907113;%%
%
\bibitem{gallex} W.~Hampel {\it et al.}, GALLEX Collaboration, 
Phys.\ Lett.\ B {\bf 447} (1999) 127.
%%CITATION = PHLTA,B447,127;%%
%
\bibitem{gno} E.~Belloti, talk at XIX International 
Conference on Neutrino Physics and Astrophysics, Sudbury, 
Canada, June 2000.
%
\bibitem{SK8B}
S.~Fukuda {\it et al.}, Super-Kamiokande Collaboration,
Phys.\ Rev.\ Lett.\  {\bf 86} (2001) 5651.
%%CITATION = HEP-EX 0103032;%%
%
\bibitem{Bahcall}
J.~N.~Bahcall, M.~H.~Pinsonneault and S.~Basu, 
Astrophys.\ J.\  {\bf 555} (2001) 990.
%%CITATION = ASTRO-PH 0010346;%%
%
\bibitem{MSW}
L.~Wolfenstein, Phys.\ Rev.\ D {\bf 17} (1978) 2369;~\ibid{20}{1979}{2634};\\
S.~P.~Mikheev and A.~Yu.~Smirnov, Sov.\ J.\ Nucl.\ Phys.\  {\bf 42} (1985) 913;
Nuovo Cim.\ C {\bf 9} (1986) 17.
%%CITATION = PHRVA,D17,2369;%%
%%CITATION = PHRVA,D20,2634;%%
%%CITATION = SJNCA,42,913;%%
%%CITATION = NUCIA,C9,17;%%
%
\bibitem{SKDN}
S.~Fukuda {\it et al.}, Super-Kamiokande Collaboration, 
Phys.\ Rev.\ Lett.\  {\bf 86} (2001) 5656.
%%CITATION = HEP-EX 0103033;%%
% 
\bibitem{fogli}
G.~L.~Fogli, E.~Lisi, D.~Montanino and A.~Palazzo, 
Phys.\ Rev.\ D {\bf 64} (2001) 093007.
%%CITATION = HEP-PH 0106247;%%
%
\bibitem{cc1}
J.~N.~Bahcall, M.~C.~Gonzalez-Garcia and C.~Pe{\~n}a-Garay,
JHEP {\bf 0108} (2001) 014.
%%CITATION = HEP-PH 0106258;%%
%
\bibitem{goswami}
A.~Bandyopadhyay, S.~Choubey, S.~Goswami and K.~Kar, 
Phys.\ Lett.\ B {\bf 519} (2001) 83.
%%CITATION = HEP-PH 0106264;%%
%
\bibitem{smirnov}
P.~I.~Krastev and A.~Yu.~Smirnov, hep-ph/0108177.
%%CITATION = HEP-PH 0108177;%%
\bibitem{NT1}
H.~B.~Nielsen and Y.~Takanishi, Nucl.\ Phys.\ B {\bf 588} (2000) 281.
%%CITATION = HEP-PH 0004137;%%
%
\bibitem{NT2}
H.~B.~Nielsen and Y.~Takanishi, Nucl.\ Phys.\ B {\bf 604} (2001) 405.
%%CITATION = HEP-PH 0011062;%%
%
\bibitem{NT3}
H.~B.~Nielsen and Y.~Takanishi, Phys.\ Lett.\ B {\bf 507} (2001) 241.
%%CITATION = HEP-PH 0101307;%%
%
\bibitem{KM}
M.~Kobayashi and T.~Maskawa, Prog.\ Theor.\ Phys.\  {\bf 49} (1973) 652.
%%CITATION = PTPKA,49,652;%%
%
\bibitem{FLN}
C.~D.~Froggatt, G.~Lowe and H.~B.~Nielsen, Nucl.\ Phys.\ B 
{\bf 414} (1994) 579.
%%CITATION = NUPHA,B414,579;%%
%
\bibitem{FNS}
C.~D.~Froggatt, H.~B.~Nielsen and D.~J.~Smith, Phys.\ Lett.\ B 
{\bf 385} (1996) 150. 
%%CITATION = HEP-PH 9607250;%%
%
\bibitem{FNGS}
C.~D.~Froggatt, M.~Gibson, H.~B.~Nielsen and D.~J.~Smith, Int.\ 
J.\ Mod.\ Phys.\ A {\bf 13} (1998) 5037.
%%CITATION = HEP-PH 9706212;%%
%
\bibitem{FF}
C.~D.~Froggatt, H.~B.~Nielsen and D.~J.~Smith, hep-ph/0108262.
%%CITATION = HEP-PH 0108262;%%
%
\bibitem{seesaw}  T.~Yanagida, in Proceedings of the Workshop on Unified
Theories and Baryon Number in the Universe, Tsukuba, Japan (1979), eds.
O.~Sawada and A.~Sugamoto, KEK Report No. 79-18; \\
M.~Gell-Mann, P.~Ramond
and R.~Slansky in Supergravity, Proceedings of the Workshop at
Stony Brook, NY (1979), eds. P.~van Nieuwenhuizen and D.~Freedman
(North-Holland, Amsterdam, 1979).  
%
\bibitem{MNS}
Z.~Maki, M.~Nakagawa and S.~Sakata, Prog.\ 
Theor.\ Phys.\  {\bf 28} (1962) 870.
%%CITATION = PTPKA,28,870;%%
%
\bibitem{Fbled}
C.~D.~Froggatt, talk at the International Workshop on ``What comes 
beyond the Standard Model?'', Bled, Slovenia, July 2001; hep-ph/0112340.
%%CITATION = HEP-PH 0112340;%%
%
\bibitem{Ramond}
P.~Ramond, talk at Les Houches EuroConference on neutrino masses 
and mixings, Les Houches, France, 18-22 June, 2001.
%
\bibitem{GJ}
H.~Georgi and C.~Jarlskog, Phys.\ Lett.\ B {\bf 86} (1979) 297;\\
J.~A.~Harvey, P.~Ramond and D.~B.~Reiss, Phys.\ Lett.\ B {\bf 92} (1980) 309.
%%CITATION = PHLTA,B86,297;%%
%%CITATION = PHLTA,B92,309;%%
%
\bibitem{Isabella}
G.~Altarelli, F.~Feruglio and I.~Masina, Phys.\ Lett.\ 
B {\bf 472} (2000) 382.
%%CITATION = HEP-PH 9907532;%%
%
\bibitem{FN}
C.~D.~Froggatt and H.~B.~Nielsen, Nucl.\ Phys.\ B {\bf 147} (1979) 277.
%%CITATION = NUPHA,B147,277;%%
%
\bibitem{Fritzsch}
H.~Fritzsch, Phys.\ Lett.\ B {\bf 70} (1977) 436.
%%CITATION = PHLTA,B70,436;%%
%
\bibitem{pierre}
H.~Arason, D.~J.~Casta{\~n}o, B.~Keszthelyi, S.~Mikaelian, 
E.~J.~Piard, P.~Ramond and B.~D.~Wright, Phys.\ Rev.\ 
D {\bf 46} (1992) 3945.
%%CITATION = PHRVA,D46,3945;%%
%
\bibitem{BC}
P.~H.~Chankowski and Z.~P{\l}uciennik, Phys.\ Lett.\ 
B {\bf 316} (1993) 312; \\
K.~S.~Babu, C.~N.~Leung and J.~Pantaleone, Phys.\ 
Lett.\ B {\bf 319} (1993) 191.
%%CITATION = HEP-PH 9306333;%%
%%CITATION = HEP-PH 9309223;%%
%
\bibitem{ADKLRCW}
S.~Antusch, M.~Drees, J.~Kersten, M.~Lindner and M.~Ratz, Phys.\ 
Lett.\ B {\bf 519} (2001) 238; \\
P.~H.~Chankowski and P.~Wasowicz, 
hep-ph/0110237.
%%CITATION = HEP-PH 0108005;%%
%%CITATION = HEP-PH 0110237;%%
%
\bibitem{SK}
Y.~Fukuda {\it et al.}, Super-Kamiokande Collaboration, 
Phys.\ Rev.\ Lett.\  {\bf 81} (1998) 1562; \\
S.~Fukuda {\it et al.}, 
Super-Kamiokande Collaboration, 
Phys.\ Rev.\ Lett.\  {\bf 85} (2000) 3999.
%%CITATION = HEP-EX 9807003;%%
%%CITATION = HEP-EX 0009001;%%
%
\bibitem{CHOOZ}
M.~Apollonio {\it et al.}, CHOOZ Collaboration, 
Phys.\ Lett.\ B {\bf 466} (1999) 415.
%%CITATION = HEP-EX 9907037;%%
%
\bibitem{cc3}
M.~C.~Gonzalez-Garcia and C.~Pe{\~n}a-Garay, private communication.
%
\bibitem{fogli2}
G.~L.~Fogli, E.~Lisi, A.~Marrone, D.~Montanino and A.~Palazzo, hep-ph/0104221.
%%CITATION = HEP-PH 0104221;%%
%
\bibitem{cc2}
M.~C.~Gonzalez-Garcia and C.~Pe{\~n}a-Garay, Phys.\ Lett.\ 
B {\bf 527} (2002) 199.
%%CITATION = HEP-PH 0111432;%%
%
\bibitem{serguey1}
S.~M.~Bilenky, D.~Nicolo and S.~T.~Petcov, hep-ph/0112216.
%%CITATION = HEP-PH 0112216;%%
%
\bibitem{cecilia}
C.~Jarlskog, Phys.\ Rev.\ Lett.\  {\bf 55} (1985) 1039.
%%CITATION = PRLTA,55,1039;%%
%
\bibitem{serguey2}
S.~M.~Bilenky, S.~Pascoli and S.~T.~Petcov, Phys.\ Rev.\ 
D {\bf 64} (2001) 053010.
%%CITATION = HEP-PH 0102265;%%
%
\bibitem{MPP}
D.~L.~Bennett and H.~B.~Nielsen, Int.\ J.\ Mod.\ Phys.\ 
A {\bf 9} (1994) 5155; \ibid{{\bf 14}}{1999}{3313}.
%%CITATION = HEP-PH 9311321;%%
%%CITATION = HEP-PH 9607278;%%
%
\bibitem{HiggsFNT}
C.~D.~Froggatt and H.~B.~Nielsen, Phys.\ Lett.\ B {\bf 368} 
(1996) 96; \\
C.~D.~Froggatt, H.~B.~Nielsen and Y.~Takanishi, 
Phys.\ Rev.\ D {\bf 64} (2001) 113014.
%%CITATION = HEP-PH 9511371;%%
%%CITATION = HEP-PH 0104161;%%

%%%%%%%%%%%%%%%%%%%%%%%%%%%%%%%%%%%%%%%%%%%%%%%%%%%%%%%
\end{thebibliography}
\end{document}